\documentclass[pre,aps,twocolumn,showpacs,floatfix]{revtex4}

\usepackage[dvips]{graphicx}
\usepackage{amsmath}
\usepackage{amssymb}

\begin{document}
\title{Transport in a L\'evy ratchet: Group velocity and distribution spread}

\author{B. Dybiec}
\email{bartek@th.if.uj.edu.pl}
\affiliation{M. Smoluchowski Institute of Physics, and Mark Kac Center for Complex Systems Research, Jagellonian University, ul. Reymonta 4, 30--059 Krak\'ow, Poland}

\author{I. M. Sokolov}
\email{igor.sokolov@physik.hu-berlin.de}
\affiliation{Institut f\"ur Physik, Humboldt-Universit\"at zu Berlin, Newtonstrasse 15, D--12489 Berlin, Germany}

\author{E. Gudowska-Nowak}
\email{gudowska@th.if.uj.edu.pl}
\affiliation{M. Smoluchowski Institute of Physics, and Mark Kac Center for Complex Systems Research, Jagellonian University, ul. Reymonta 4, 30--059 Krak\'ow, Poland}

\date{\today}
\begin{abstract}
We consider the motion of an overdamped particle in a periodic
potential lacking spatial symmetry under the influence of
symmetric L\'evy noise, being a minimal setup for a ``L\'evy
ratchet.'' Due to the non-thermal character of the L\'evy noise,
the particle exhibits a motion with a preferred direction even in
the absence of whatever additional time-dependent forces. The
examination of the L\'evy ratchet has to be based on the
characteristics of directionality which are different from
typically used measures like mean current and the dispersion of
particles' positions, since these get inappropriate when the
moments of the noise diverge. To overcome this problem, we discuss
robust measures of directionality of transport like the position
of the median of the particles displacements' distribution
characterizing the group velocity, and the interquantile distance
giving the measure of the distributions' width. Moreover, we
analyze the behavior of splitting probabilities for leaving an
interval of a given length unveiling qualitative differences
between the noises with L\'evy indices below and above unity.
Finally, we inspect the problem of the first escape from an
interval of given length revealing independence of exit times on
the structure of the potential.
\end{abstract}

\pacs{
 05.40.Fb, 
 05.10.Gg, 
 02.50.-r, 
 02.50.Ey, 
 }
\maketitle

%
%
\section{Introduction\label{sec:introduction}}

Motion of particles in an external periodic potential with broken
spatial symmetry and under influence of external forces with zero
mean can result in occurrence of the persistent, directed current
\cite{astumian1994,magnasco1994,kula1998,reimann2002,reimann2002b}.
The occurrence of the directed transport is a consequence of the
violation of the detailed balance in systems acting away from
equilibrium. Typically, one assumes that the overall force acting
on the particle is a superposition of the Gaussian thermal noise
with another periodic or stochastic force \cite{reimann2002}. In
those cases all moments of the distribution of the noisy force do
exist, and the existence of moments of
the distribution of the particle's velocity is guaranteed. There are at least
two reasons to consider the directed motion of a particle in a
periodic ratchet potential under the influence of a heavy-tailed
noise causing anomalously large particle displacements. One of
them is pursuing the line of investigations of rectification of
non-thermal noises and generalizing the corresponding
considerations to the heavy-tailed cases which were found to be
quite abundant under non-equilibrium conditions
\cite{shlesinger1995,nielsen2001,dubkov2005b,dubkov2007}. Another
one, connected with the former, is seeking for the way to
characterize the ensuing directed motion in the case when the
corresponding statistical moments are absent, so that neither the
dispersion nor even the mean of the corresponding displacements
exist. In consequence, in such cases the standard characteristics
of motion like mean velocity or Peclet number become inapplicable.

The interplay of deterministic dynamics and perturbative
L\'evy-type noises have been addressed in literature in various
scenarios including several noise-induced effects like resonant
activation \cite{dybiec2004,dybiec2004b}, stochastic resonance
\cite{dybiec2006b}, dynamical hysteresis
\cite{dybiec2006b,dybiec2007e}, studies of decay/relaxation
properties of the probability densities
\cite{dybiecphd,chechkin2004}, escape from bounded intervals
\cite{dybiec2006,zoia2007}, the classical barrier crossing problem
\cite{ditlevsen1999,chechkin2005,dybiec2007} or examination of
stationary states
\cite{chechkin2002,chechkin2003,chechkin2004,dybiec2007d}.
However, very few examples \cite{delcastillonegrete2007} tackle
the problem of L\'evy noise driven dynamics of periodic systems.

In the present work we study the behavior of a particle in the
periodic potential with a broken spatial symmetry subjected to the
action of a symmetric L\'evy stable noise. 
The $\alpha$-stable noise can 
be expected to occur in systems being out of thermal equilibrium, 
where detailed balance conditions is violated. In such realms one 
can anticipate noise induced current in a static potential with broken
spatial symmetry. Consequently, what we address here as the L\'evy
ratchet resembles so called thermal ratchet \cite{reimann2002}
with time-dependent temperature variations. Heavy tails of distribution of
additive stochastic increments lead however to considerable
peculiarities of such motion, due to power-law character of
probability densities of the particle's displacements, which lack
the dispersion and may also lack the mean. Therefore, the
examination of the current defined as the time derivative of the
mean position may not be adequate for the L\'evy ratchet, and
one has to look for other quantities which characterize
the motion caused by the interplay of the noise and a potential.
Instead of examination of the current and of the effective
diffusion coefficient we discuss several robust probabilistic
characteristics of the particle's displacement based on its
cumulative distribution: the behavior of median, which allows to
introduce the group velocity, the growth of interquantile
distances and the behavior of splitting probabilities (as defined in Sec.~\ref{sec:split}).

As it will be discussed futher on,
symmetric stable noise acting in a presence of a potential with broken
spatial symmetry can induce directed motion of a test particle in an overdamped system.
As long as the mean
value of the perturbative noise exists and equals zero the displacement of the
particle also possesses a mean. The overall motion can be then
characterized by the temporal changes in the mean position, or, in the
other words, by a mean velocity. In the case of an ensemble of
particles mean velocity corresponds to the overall current,
which is an adequate measure of the directionality of motion
\cite{reimann2002}. This situation is realized for $\alpha$-stable
noises with $1<\alpha<2$.

If the noise, however, is distributed according to a L\'evy-stable law with
$0<\alpha <1$, neither the mean of the noise nor the mean of the overall
displacement exist, so that one has to look for other quantities characterizing
the directed motion. The discussion of these quantities, parallelly to
investigation of the properties of the directed transport, is the main topic of
the present article. Its structure is as follows. The next section
(Section~\ref{sec:model}) presents the model under consideration. The obtained
numerical results are included in Section~\ref{sec:results}, which is devoted to
various characteristics of the considered ratcheting device. The cumulative
distribution of the overall displacement probability is examined in
Section~\ref{sec:cdf}, whereas its main features like median allowing for a
definition of a group velocity of the probability packet are studied in
Section~\ref{sec:median}. We discuss also interquantile distance
(Section~\ref{sec:interquantile}), as well as splitting probability
(Section~\ref{sec:split}) that defines the fraction of current going to the left
versus right direction when leaving an interval of finite length. Finally,
(Sec.~\ref{sec:et}) we discuss properties of the first escape time
distribution from a box of given width. The paper is closed with summary and
concluding remarks (Section~\ref{sec:summary}).

%
%
\section{Model\label{sec:model}}

\begin{figure}[!ht]
\begin{center}
\includegraphics[angle=0,width=8.0cm]{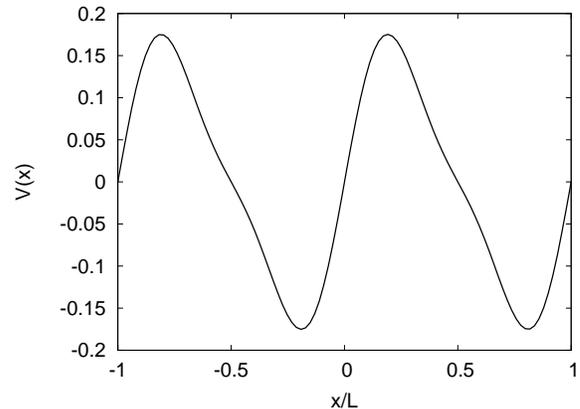}
\caption{The potential, $V(x)$, with broken spatial symmetry given by Eq.~(\ref{eq:potential}) used for examination of the studied L\'evy ratchet.}
\label{fig:potential}
\end{center}
\end{figure}

In what follows we consider the system described by the 
overdamped Langevin equation
\begin{equation}
\frac{dx}{dt}=-V'(x)+\zeta(t),
\label{eq:model}
\end{equation}
where $\zeta(t)$ stands for a L\'evy stable noise, i.e. the noise for which
increments are distributed according to a stable probability
density. Finally, $V(x)$ is an external static ratchet potential, see
Fig.~\ref{fig:potential}, with broken spatial symmetry
\begin{equation}
V(x)=\frac{1}{2\pi}\left[ \sin\frac{2\pi x}{L} +\frac{1}{4}\sin\frac{4\pi x}{L} \right].
\label{eq:potential}
\end{equation}
Initially, at $t=0$, a test particle is located at $x=0$.

%
%

In the most general case L\'evy distributions
correspond to a 4-parameters
family of the probability density functions \cite{feller1968,janicki1994,janicki1996}.
Here we restrict studies to symmetric stable noises, $L_{\alpha}(\zeta;\sigma,\mu)$, which are characterized by the
Fourier-transform $\phi(k) = \int_{-\infty}^\infty e^{ik\zeta}
L_{\alpha}(\zeta;\sigma,\mu) d\zeta$ of the probability density given by
\cite{feller1968,janicki1994,janicki1996}
\begin{equation}
\phi(k) = \exp\left[ ik\mu -\sigma^\alpha|k|^\alpha \right].
\label{eq:charakt}
\end{equation}
The parameter $\alpha$ (where $\alpha\in(0,2]$) is the stability index of
the distribution describing (for $\alpha <2$) its asymptotic ``fat''
tail characteristics yielding $L_{\alpha}(\zeta; \sigma,\mu)\sim
|\zeta|^{-(1+\alpha)}$ for large $\zeta$. The parameter $\mu$ denotes the location
parameter representing position of the modal value. The Gaussian distribution, see Eq.~(\ref{eq:charakt}),
corresponds to a special case of a L\'evy law for $\alpha=2$, with
$\mu$ interpreted now as a mean and $\sigma$ as the dispersion of the
distribution.
Additionally, we assume $\mu=0$ and $\sigma=\{0.25,0.5,1\}$, i.e. we consider
symmetric, strictly stable noises only.
It is worthy mentioning that nonzero asymmetry parameter, which is not considered here, introduces a
preferred direction of noise pulses and consequently may lead to a preferred
direction of motion \cite{dybiec2007d,delcastillonegrete2007}.

%
%
\section{Results\label{sec:results}}

For symmetric L\'evy noises, Eq.~(\ref{eq:model}) is equivalent to the following fractional differential Fokker-Planck equation \cite{metzler1999,yanovsky2000,schertzer2001}
\begin{equation}
\frac{\partial P(x,t)}{\partial t} = \frac{\partial}{\partial x}V'(x)P(x,t)+\sigma^\alpha\frac{\partial^\alpha P(x,t)}{\partial |x|^\alpha},
\label{eq:ffpe}
\end{equation}
where $\alpha$ and $\sigma$ characterize the noise.
The fractional (Riesz-Weyl) derivative is interpreted in
the sense of the Fourier transform
\cite{jespersen1999,chechkin2002,chechkin2003}
$\frac{\partial^\alpha}{\partial|x|^\alpha}f(x)=-\int_{-\infty}^{\infty}\frac{dk}{2\pi}e^{-ikx}|k|^\alpha
\hat{f}(k)$. Nevertheless, due to possible instabilities \cite{meerschaert2004} of numerical approximations to Eq.~(\ref{eq:ffpe}) \cite{gorenflo2002} in the following studies we use approach based solely on the Langevin equation~(\ref{eq:model}).

The results presented below were obtained by the numerical integration
of Eq.~(\ref{eq:model}) with L\'evy stable noises using the methods discussed
in Refs. \cite{janicki1994,janicki1996,dybiec2006,dybiec2007}. The
numerical integration of Eq.~(\ref{eq:model}) was performed with the
time step of $\Delta t=10^{-3}$. The number of realizations
varied from $N=10^5$ to $N=10^6$ leading in all cases to consistent results.

Initially, a test particle is located at $x=0$. In course of time the
distribution of the particle's positions $x$ evolves due to the presence of the
stochastic force and of the deterministic potential in Eq.~(\ref{eq:model}). As
we proceed to show, the ensuing coarse-grained probability distribution of $x$
is a symmetric distribution of the power-law type (Sec.~\ref{sec:cdf}). The
overall motion of the probability density is characterized by changes in the
location of its median (Sec.~\ref{sec:median}). The width of the probability
density grows with time. The growing width of the distribution is characterized
by growing interquantile distance (Sec.~\ref{sec:interquantile}). Still another
characteristic of the motion can be addressed by analyzing behavior of splitting
probabilities: A particle initially located at $x=0$ can escape from an interval
of given width $L$ centered at the initial position to the left or to the right.
The statistics of the corresponding escape events (splitting probabilities,
Sec.~\ref{sec:split}) provides another measure of the directionality of motion. Finally,
mean first escape time provides a characteristic of the time scales involved in
such escapes (Sec.~\ref{sec:et}).

%
%
\subsection{Cumulative distribution\label{sec:cdf}} 
\begin{figure}[!ht]
\begin{center}
\includegraphics[angle=0,width=8.0cm]{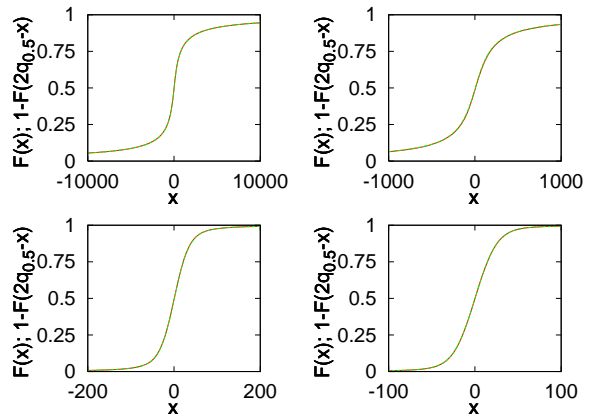}
\caption{(Color online) Cumulative distributions $F(x,t=100)$ and
$1-F(2q_{0.5}-x,t=100)$ for $\alpha=0.7$ (top left panel), $\alpha=0.9$ (top
right panel), $\alpha=1.5$ (bottom left panel), $\alpha=1.7$ (bottom
right panel) with $\sigma=1$. The perfect agreement between both curves indicates the symmetric shape of corresponding PDFs with respect to medians. The same agreement is observed for smaller values of the noise intensity $\sigma$. Note the large difference in scales for different
values of $\alpha$.
}
\label{fig:cdf}
\end{center}
\end{figure}

We first discuss
the cumulative distribution function (CDF) of the particle's displacement
under the non-equilibrium additive noise. The CDF is given by $F(x,t) = \int_{-\infty}^x P(x',t)dx'$ with $P(x,t)$ being
the corresponding probability density function (PDF), which can be interpreted in terms of particle concentration. Simulations show, that in course of time the corresponding PDF attains a symmetric form, see Fig.~\ref{fig:cdf}. The
 density $P(x,t)$ is then found to move in one direction and to
broaden in time. These effects can be seen by inspection of the overall position of the particles
 characterized by the median of the distribution $q_{0.5}$ (cf. Fig.~\ref{fig:median})
and its width defined by the interquantile distance, e.g. $q_{0.8}(t)-q_{0.2}(t)$ (cf. Fig.~\ref{fig:interquantile}).
The quantiles $q_m$ ($0<m<1$) are defined according to the relation $F(q_m,t) = m$.

For symmetric probability densities, i.e. densities such that $P(q_{0.5}-x,t)=P(x+q_{0.5},t)$
the cumulative density $F(x,t)$ fulfills
\begin{equation}
F(x,t)=1-F(2q_{0.5}-x,t).
\end{equation}
For the L\'evy ratchet system, the constructed CDFs prove that
 PDF recorded at given time are symmetric with respect to median ($q_{0.5}$) if
the space scales much larger than the period of the potential are
considered, see Fig.~\ref{fig:potential} and~\ref{fig:cdf}.

%
%

\begin{figure}[!ht]
\begin{center}
\includegraphics[angle=0,width=8.0cm]{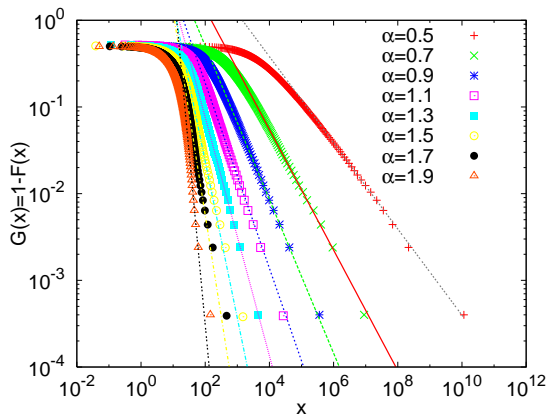}
\caption{(Color online) `Survival probability' $G(x,t=100)=1-F(x,t=100)$ with $\sigma=1$. For
large $x$, numerical simulations demonstrate power-law character of
`survival probabilities' $G(x,t=100)$. Nevertheless, the value of the exponent
characterizing large $x$ behavior is different from $\alpha$ when
underlying noise is characterized by $\alpha\geqslant1.3$, see
Fig.~\ref{fig:survivalexponent}}
\label{fig:survival}
\end{center}
\end{figure}

\begin{figure}[!ht]
\begin{center}
\includegraphics[angle=0,width=5.0cm]{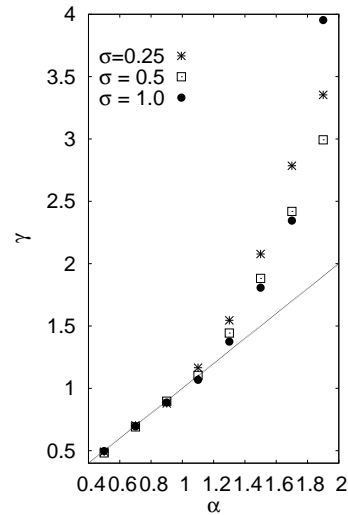}
\caption{Value of the exponent $\gamma$ in `survival probability'  describing the
asymptotic behavior of its tails $G(x,t)\propto x^\gamma$, see Fig.~\ref{fig:survival}. For
$\alpha\geqslant 1.3$, the exponent characterizing tails' asymptotic differs
from the index $\alpha$ of the underlying noise because the motion of the test particle starts to be more sensitive to the structure of the potential. Relative errors of the fit are
smaller than 1\% of value of exponents. Consequently, error bars are smaller
than symbol size.}
\label{fig:survivalexponent}
\end{center}
\end{figure}

The distributions of the particles' displacements in
Fig.~\ref{fig:cdf} exhibit power-law tails. To detect them it is
enough to consider the asymptotic behavior of the `survival
probability' $G(x,t)=1-F(x,t)$ for large $x$, see
Fig.~\ref{fig:survival} clearly displaying this power-law behavior.
The simplest hypothesis here would be that the corresponding
exponent is the same as the one of the noise; if $\zeta(t)$ has a
stable density characterized by exponent $\alpha$, i.e. $P(\zeta,t)
\propto \zeta^{-(1+\alpha)}$ for large $\zeta$, the `survival
probability' should behave like $G(x,t) \propto x^\gamma$ with
$\gamma = \alpha$. This kind of behavior is indeed observed for
the values of stability index $\alpha$ smaller or around one, see
Fig.~\ref{fig:survivalexponent}. For larger $\alpha$, $G(x,t)$ still
shows a power-law asymptotic, see Fig.~\ref{fig:survival},
however, the numerically obtained values of exponents $\gamma$ differ from
$\alpha$, see Fig.~\ref{fig:survivalexponent}. This difference is
pertinent to a very far tail of the distribution, since the
behavior of the interquantile distances, say
$q_{0.2}(t)-q_{0.8}(t)$, unveils the behavior compatible with the
$\gamma = \alpha$ assumption for all $\alpha$, see
Sec.~\ref{sec:interquantile}.

Inspection of `survival probability' $G(x,t)$, see
Fig.~\ref{fig:survival}, suggests that for small value of $\alpha$ the
particle practically does not feel the potential. Therefore, the particle behaves
like a free particle and the exponent characterizing `survival
probability' is the same as the one characterizing statistical
properties of the noise. This observation is in accordance with
predictions of the continuous time random walks theory in situations
when the mean waiting time for a next jump is finite \cite{metzler2000,metzler2004}. 
For increasing
values of $\alpha$, the motion gets to be more sensitive to the structure of
the potential. Consequently, the form of $G(x,t)$  
in the intermediate range of $x$ diverges from the pure L\'evy distribution. 
The difference between $\gamma$ and $\alpha$ is probably explained
by extremely slow convergence of the tail of the corresponding distribution
to its asymptotic form.

In order to fit the tail asymptotics properly, it is necessary to have large
statistics corresponding to extreme events which should cover at least a couple
of orders of magnitude. Such a coverage is rather easily reached for values of
$\alpha\leqslant 1.1$. However, for large values of the stability index $\alpha$
($\alpha>1.1$), the statistics of `survival probability' $G(x,t)$ for very large
values of $x$ is quite poor: the
exploration of large $x$ dependence requires simulation times which are beyond
our possibilities. The values of $\gamma$ obtained in simulations did not
indicate  changes for chosen simulation times, however the
fitted slope was found to be sensitive to the choice of the threshold from which
a power-law is fitted.

%
%

%
%

%
%
\subsection{Median and group velocity\label{sec:median}}
\begin{figure}[!ht]
\begin{center}
\includegraphics[angle=0,width=8.0cm]{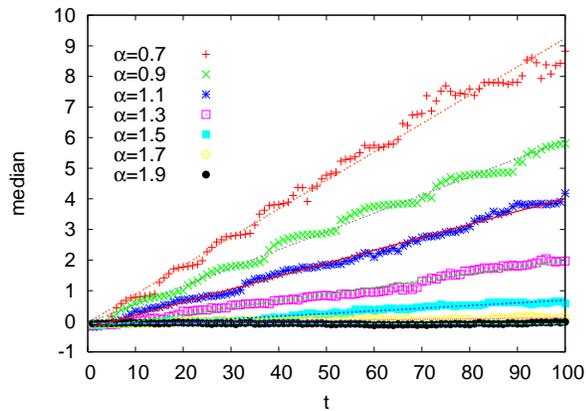}
\caption{(Color online) Time dependence of the location of median,
$q_{0.5}(t)$ which describes the overall motion of the ensemble of
particles injected into the L\'evy ratchet at $t=0$ ($\sigma=1$).
The linear growth of median with time allows to define the group velocity
of the packet of particles.}
\label{fig:median}
\end{center}
\end{figure}

\begin{figure}[!ht]
\begin{center}
\includegraphics[angle=0,width=8.0cm]{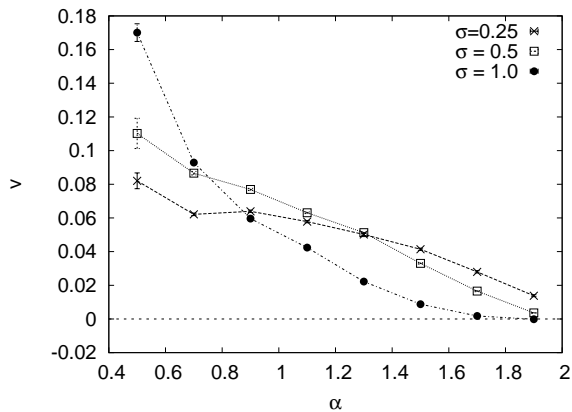}
\caption{Value of the group velocity (coefficient $v$ in
Eq.~(\ref{eq:median}) characterizing slope of curves in
Fig.~\ref{fig:median}) as a function of the stability index
$\alpha$. The lines are drawn to guide the eye only. The fastest overall
motion of the probability density is observed for small values of
stability index $\alpha$ for which large noise increments are more
probable. The value of $v$ vanishes towards $\alpha=2$ corresponding
to the equilibrium Gaussian noise for which persistent current disappears.
Error bars represent errors of the fit. Error bars, except for those for $\alpha=0.5$, are smaller than symbol size.
}
\label{fig:slope}
\end{center}
\end{figure}

The position of the median describes the overall motion of the
probability density to find a particle in the vicinity of $x$.
In our case, the corresponding PDF coarse-grained over the period of the
potential, is a symmetric, monomodal function, so that the
temporal change of the median (which coincides with the maximum of the
PDF) gives us the possibility to define the group velocity of the
particles' packet as $v = dq_{0.5}/dt$. Our numerical simulations show that the
position of median $q_{0.5}(t)$ changes linearly with time
\begin{equation}
q_{0.5}(t)=v\cdot t+b,
\label{eq:median}
\end{equation}
which corresponds to the constant group velocity. Note that this group
velocity can be defined even in the case when the mean velocity or
current does not exist due to large fluctuations, and defines the
behavior of the typical displacement of particles with
time. Fig.~\ref{fig:median} presents location of the median as a
function of time for different values of stability exponents
$\alpha$. The `periodic-like' modulation of the median position is
caused by the periodic shape of the potential, see ordinate of
Fig.~\ref{fig:potential}. For the constant amplitude of the noise the
group velocity decreases with increase of the exponent $\alpha$ and
disappears in the Gaussian case $\alpha =2$, see Fig.~\ref{fig:slope},
in which case the noise satisfies the detailed balance condition.

%
%

The `survival probability' shows that asymptotic behavior of the probability
density of a particle position is not sensitive to the detailed structure of the
potential. The presence of potential is visible for larger values of $\alpha$
when the exponent characterizing the tails of distributions starts to differ
from $\alpha$, see Figs.~\ref{fig:survival} and \ref{fig:survivalexponent}.
Different situation takes place for median which reflects the structure of the potential. 
It is manifested in the periodic modulation
of median position visible in Fig.~\ref{fig:median}. This periodic structure has
the same period like the potential. This suggests that tails of distributions can
freely overpass the potential maxima, while the motion of median is strongly
affected by presence of maxima of the potential.

%
%

%
%
\subsection{Interquantile distance\label{sec:interquantile}}

\begin{figure}[!ht]
\begin{center}
\includegraphics[angle=0,width=8.0cm]{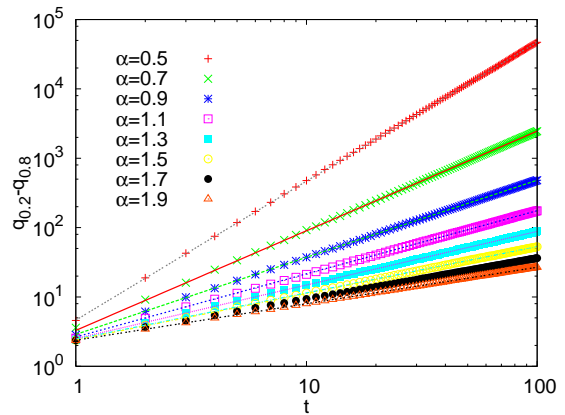}
\caption{(Color online) Time dependence of the interquantile distance
$q_{0.2}(t)-q_{0.8}(t)$ with $\sigma=1$ together with fitted theoretical curves.
The decrease of scale parameter $\sigma$ makes the particle more sensitive to the shape of the potential and consequently introduces periodic-like modulation of interquantile distances (results not shown).}
\label{fig:interquantile}
\end{center}
\end{figure}

\begin{figure}[!ht]
\begin{center}
\includegraphics[angle=0,width=8.0cm]{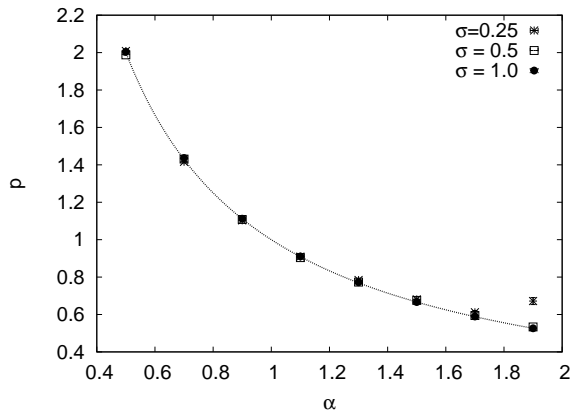}
\caption{Value of the exponent $p$ in Eq.~(\ref{eq:interquantile}) as a
function of the stability index $\alpha$. Line presents theoretical
dependence of the exponent characterizing growth of the interquantile
distance, i.e. $1/\alpha$. Error bars are smaller than symbol size.}
\label{fig:interquantileexponent}
\end{center}
\end{figure}

Interquantile distance is a robust measure characterizing the
distribution's width even in cases when moments of the distribution do not exist.
For the studied L\'evy ratchet
this interquantile distance scales like $t^{p}$ independently of the value of the noise intensity $\sigma$, i.e.
\begin{equation}
q_m(t)-q_{1-m}(t) \propto t^{p},
\label{eq:interquantile}
\end{equation}
where $p=1/\alpha$.
Fig.~\ref{fig:interquantile} presents time dependence of interquantile
distance $q_{0.2}(t)-q_{0.8}(t)$ along with fitted
curves. Fig.~\ref{fig:interquantileexponent} presents values of fitted
exponents, together with the predicted $1/\alpha$-behavior. The interquantile
distance $q_m(t)-q_{1-m}(t)$ scales in the same manner also for other values of $m$ ($0<m<1$).
Note that the periodic modulation which is typical for evolution of median,
see Fig.~\ref{fig:median}, can be not visible for interquantile distance.
Essentially, this modulation does not appear in the behavior of whatever
quantiles $q_{m}(t)$ with $m\neq0.5$ and sufficiently large noise intensity $\sigma$.
The decrease of the noise intensity $\sigma$, introduces periodic modulation also to interquantile distance which can be the most easily observed for large values of the stability index $\alpha$ (figure not shown).

%
%

%
%

%
%
\subsection{Splitting probability\label{sec:split}}

\begin{figure}[!ht]
\begin{center}
\includegraphics[angle=0,width=8.0cm]{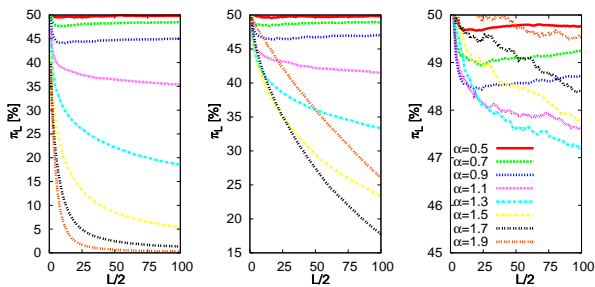}
\caption{(Color online) Splitting probability, $\pi_L$, i.e. probability of the
first escape to the left as a function of the box
half-width. Initially particle is located in the middle of the box of
half-width $L/2$, see Eq.~(\ref{eq:potential}) and
Fig.~\ref{fig:potential}. Various panels correspond to different values of the scale parameter $\sigma$: $\sigma=0.25$ (left panel), $\sigma=0.5$ (middle panel) and $\sigma=1$ (right panel).}
\label{fig:split}
\end{center}
\end{figure}

Considering splitting probabilities gives us another characteristics
of the directionality of motion. Splitting probability is a measure of the preferred
direction to which the first escape from the system takes
place. In this case we consider an interval of length $L$
centered around the point where particles are introduced, and consider
the probability for the particle to escape the interval
through its left (right) boundary.
Fig.~\ref{fig:split} presents probability $\pi_L$ of the first escape
to left as a function of the box half-width. The presented splitting
probability indicates that the preferred direction of the first escape
from the system is to the right, since for longer intervals $\pi_L < 1/2$.
Furthermore, decrease of the scale parameter $\sigma$ makes this tendency stronger.

%
%

Fig.~\ref{fig:split} shows a clear difference between the
cases $1 < \alpha < 2$ and $\alpha < 1$. In the first case for the
intervals (``boxes'') much larger than the period of the potential the
probability $\pi_L$ decays monotonically with the box size, which
means than more and more particles leave the box through its right
boundary. In the second case, as exemplified by $\alpha = 0.9$ the
behavior of $\pi_L$ is nonmomotonical: it decays with $L$ up to $L/2
\approx 20$ and then starts to grow again.

This striking difference between the results for $\alpha>1$ and
ones for $\alpha < 1$ gets clear when returning to the behavior of
the interquantile distances. In the first case the shift of the
distribution's median (coinciding with its maximum), going linearly in
time, is asymptotically faster than the growth of the distribution's
width, which allows us to discuss the position of the typical particle
as a mean position and to introduce the mean current. In course of the time
(necessary to leave the larger box) the typical displacement wins
over the spread, and most of the particles come out of the box
through its one (right) boundary.

In the case $\alpha < 1$ the width of the distribution grows faster than its
maximum moves. Due to the well-prescribed initial position, the contribution
of the overall motion is still perceptible at shorter times and thus for smaller
boxes, and the difference between the probabilities to leave the box through
its left and right boundary grows. At longer times the broadening of the
distribution prevails, typical
displacement gets hardly relevant on the background of
fluctuations, and the difference in splitting probabilities decays.
However, the distribution still has a pronounced
maximum, and its motion defines the group velocity.

%
%
\subsection{Exit Time\label{sec:et}}

The spiting probability is a measure of directionality of the first escape from
a box of a given width. The quantity which is very closely related to the
spiting probability is exit time, i.e. the time which is needed to escape from
the same box. Here we study escape time, $\tau$, which is the first escape time
from the box through one of its edges, i.e. we do not distinguish between
borders of the box.

\begin{figure}[!ht]
\begin{center}
\includegraphics[angle=0,width=8.0cm]{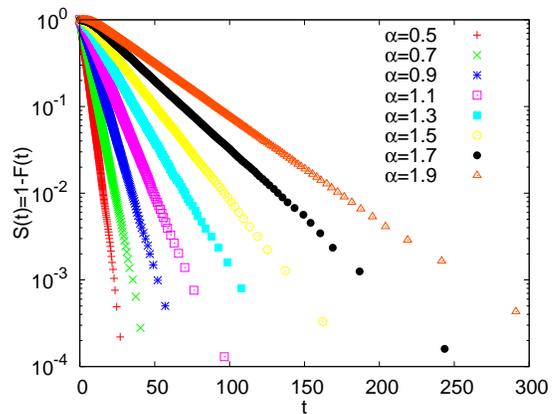}
\caption{(Color online) Survival probability, $S(t)=1-F(t)$, i.e. the probability of finding a particle in the box of the width $L=10$ for $\sigma=1$. Survival probabilities are of the exponential type regardless of the value of $L$ and $\sigma$.}
\label{fig:etd}
\end{center}
\end{figure}

The same information which is included in the escape time distribution is
included in the survival probability. Survival probability, $S(t)=1-F(t)$, is the
probability of finding a particle in the box of given width at time $t$, i.e. it
is the ratio of particles which at time $t$ are still in the box. The survival
probability $S(t)$ is exponential function of time
\cite{dybiec2006,dybiec2007,imkeller2006}. In the Fig.~\ref{fig:etd}, survival
probabilities for $L=10$ are presented with $\sigma=1$. Results for other values of the box width
$L$ and scale parameter $\sigma$ also show exponential behavior.

%
%

%
%
\begin{figure}[!ht]
\begin{center}
\includegraphics[angle=0,width=8.0cm]{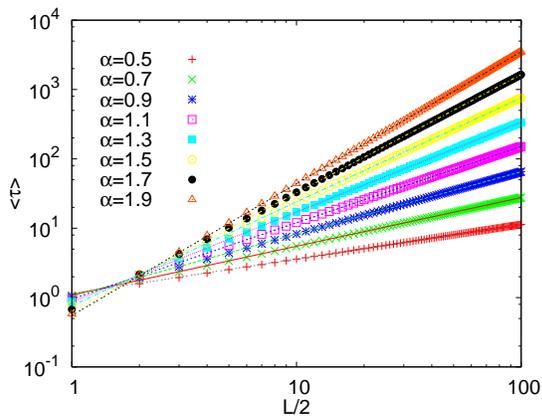}
\caption{(Color online) Mean exit time, $\langle\tau\rangle$, i.e. the average of the
first escape time from the box as a function of the box
half-width for $\sigma=1$. Initially a particle is located in the middle of the box of
the half-width $L/2$, see Eq.~(\ref{eq:potential}) and
Fig.~\ref{fig:potential}.}
\label{fig:met}
\end{center}
\end{figure}

\begin{figure}[!ht]
\begin{center}
\includegraphics[angle=0,width=8.0cm]{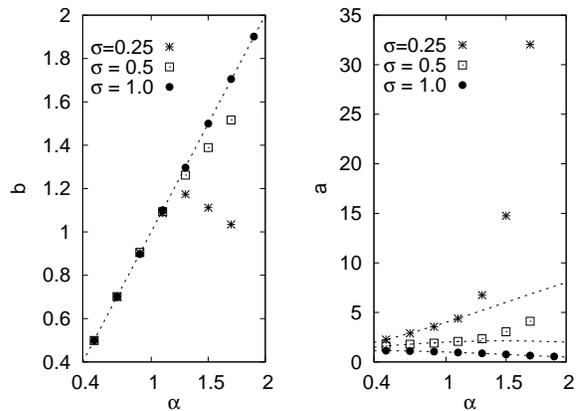}
\caption{Value of the exponent $b$ and the prefactor $a$ in Eq.~(\ref{eq:met}) as a
function of the stability index $\alpha$. Lines present theoretical
values of the exponent and the prefactor characterizing mean escape time
for a free particle, i.e. $\alpha$ and $1/[\Gamma(\alpha+1)\sigma^\alpha]$ respectively.}
\label{fig:metexponent}
\end{center}
\end{figure}

From first escape time distribution it is possible to calculate the mean exit time. For a free particle starting its motion in the middle of the interval of width $L$ the mean exit time is \cite{zoia2007}
\begin{equation}
\langle\tau(x(0)=0)\rangle = \frac{(L/2)^\alpha}{\sigma^\alpha\Gamma(\alpha+1)}=a\cdot\left(\frac{L}{2}\right)^{b}.
\label{eq:met}
\end{equation}
From Eq.~(\ref{eq:met}) it implies that $\langle \tau \rangle \propto L^\alpha$.
The same kind of behavior can be observed for the L\'evy ratchet.
For the large value of the scale parameter $\sigma$, the presence of the potential does not affect the
dynamics of the particle. Consequently, both the prefactor and exponent are like in the free case, see Eq.~(\ref{eq:met}). The decrease of the noise intensity affects
$L^\alpha$ scaling and the prefactor in Eq.~(\ref{eq:met}),
especially in situations when the L\'evy noise is close to the Gaussian one
and small values of the scale parameter $\sigma$ are considered.
Nevertheless, for small to moderate values of the stability index $\alpha$, with all considered values of noise intensity $\sigma$, the first escape time is like in the free case.
Mean exit time as a
function of the box half-width with $\sigma=1$ is depicted in Fig.~\ref{fig:met}.
Fig.~\ref{fig:metexponent} tests applicability of Eq.~(\ref{eq:met}).

%
%

%
%
\section{Summary and Conclusions\label{sec:summary}}

Alltogether, we have numerically documented, that 
the motion of the overdamped particle subjected to a symmetric
L\'evy noise in a potential with a broken spatial symmetry leads
to occurrence of directionality of motion. The preferred
direction of motion results as a consequence of an interplay
between non equilibrium character of the underlying stable noise
and the broken spatial symmetry of the static potential. However,
the characteristics of the particle's displacement, like mean
velocity and the diffusion coefficient, describing the particle's
dispersion, get inappropriate when white L\'evy noises are
considered. Diverging variance and possibly non-existing mean make
it necessary to introduce other characteristics of the directed
motion. Consequently, the conducted research is based on robust
measures characterizing the collective motion of L\'evy-Brownian
particles, like group velocity, interquantile width of the
displacements' distribution and splitting probabilities.

In the system under the study, the initially sharp distribution of
particles' positions broadens in course of time leading to
symmetric probability density of the power-law type. The median of
the distribution, which characterizes the group velocity of the
particle packet, moves linearly with time. The fastest motion of
median is observed for small values of the stability exponent
$\alpha$. The coherence of the motion can be characterized by the
interquantile width of the distribution of the particles'
position, whose growth with time was found to follows the
$q_m(t)-q_{1-m}(t) \propto t^{1/\alpha}$ pattern. For $\alpha > 1$
the width of the distribution grows therefore slower than its
maximum moves, contrary to the situation when $0<\alpha<1$ and
the opposite is the case. This fact is mirrored in the behavior of
splitting probabilities for leaving an interval of a given length.

Eq.~(\ref{eq:model}) describes the dynamics of the L\'evy ratchet
for any value of the noise parameters. Here, we restricted our
studies to symmetric noises only. In such a case the only relevant
noise parameter is the stability index $\alpha$ which
characterizes the asymptotic power law behavior of additive noise
pulses. The stability index controls the level of non-equilibrity
of the noise. On the one hand, for values of $\alpha$ smaller than
2, the L\'evy noise serves as a prototypical perturbation source
that drives the system out of equilibrium. On the other hand, for
$\alpha=2$, the L\'evy noise becomes Gaussian noise and models
thermal fluctuations in the vicinity of the equilibrium state. The
rectification, ratcheting effect is visible only for values of
$\alpha<2$. The increase of the stability index $\alpha$ decreases
the ratcheting effect. Finally for $\alpha=2$, similarly to the
original ratchet and pawl device \cite{reimann2002}, the absence
of the average particles' current is a simple consequence of the
second law of thermodynamics: despite the broken spatial symmetry
assured by the form of the potential, no systematic preferential
motion of the random dynamics can be detected and the ratcheting
effect disappears.

The emergence of the rectification in the ``minimal L\'evy ratchet''
(Eq. (1)) follows as a result of the non-equilibrium type of the stochastic
driving and broken spatial symmetry of the potential. On the one hand, the
ratcheting effect is observable for symmetric noises, therefore it is induced by
the potential. On the other hand, in the minimal L\'evy ratchet scenario the presence of
the potential is not always explicit: For a free L\'evy particle the probability
density of particle positions is given by a L\'evy distribution with growing
width and the same tails characteristics as the underlying noise. The same
effect is observed here for the L\'evy ratchet with $\alpha<1.3$ when the tail
asymptotic of the PDF for a particle position becomes characterized by the same
exponent like the underlying noise. For $\alpha \geqslant 1.3$ these exponents
start to differ suggesting that only for $\alpha$s excessing a certain threshold
value the particle starts to feel the potential. On the contrary, closer
examination of the median reveals periodic structure of the median position with
the same space period as the one characterizing the potential. Finally, for all
values of the stability index $\alpha$ with sufficiently large noise intensity
$\sigma$, values of the mean exit time scale as in the free particle case. The
decrease of the scale parameter $\sigma$ makes the system more sensitive to the presence of a potential.

For a L\'evy ratchet, symmetric stable noise together with
periodic potential with broken spatial symmetry provide a minimal
setup for the occurrence of directed current. Contrary to Gaussian
noises, stable L\'evy ones can be intrinsically asymmetric; the
asymmetry of the noise introduce preferred direction of the noise
pulses might affect the current. On the other hand, the asymmetry
of noise breaks symmetry of the system and introduces a component
which is of `non-zero average type.' The asymmetric L\'evy noise
can be still incorporated to the L\'evy ratchet in the
`zero-average manner.' This can be reached by making asymmetry
parameter a periodic function of time \cite{dybiec2007e}. The time
dependence of asymmetry parameter will slightly modify the type of
the L\'evy ratchet making it closer to the ones of either
`thermal' or `tilting' type. Considering such situations might be
an interesting further step pursuing our line of investigations.

%
%
\begin{acknowledgments}
Authors acknowledge stimulating and inspiring
discussions with E. Barkai and I. Pavlyukevich.
The research has been supported by the Marie Curie TOK COCOS grant (6th EU
Framework Program under Contract No. MTKD-CT-2004-517186) and European Science
Foundation (ESF) via `Stochastic Dynamics: fundamentals and applications'
(STOCHDYN) program. \ Additionally, BD acknowledges the support from the
Foundation for Polish Science and the hospitality of the Humboldt University of
Berlin and the Niels Bohr Institute (Copenhagen). The support by DFG within
SFB555 is also acknowledged.
\end{acknowledgments}

%
%

\end{document}